\documentclass[preprint,12pt]{elsarticle}

\usepackage{graphics}

\usepackage{amssymb}
\usepackage{bm}

\usepackage{amsthm}

\begin{document}

\begin{frontmatter}



\title{Fidelity Approach in Topological Superconductors with Disorders}

\author{Wen-Chuan Tian}
\author{Guang-Yao Huang}
\author{Zhi Wang \corref{cor1}}
\author{Dao-Xin Yao \corref{cor2} }
\cortext[cor1]{Corresponding author: Zhi Wang; Email: physicswangzhi@gmail.com; Tel: +86-20-84111107}
\cortext[cor2]{Corresponding author: Dao-Xin Yao; Email: yaodaox@mail.sysu.edu.cn; Tel: +86-20-84112078}
\address{School of Physics and Engineering, Sun Yat-sen University, Guangzhou 510275, China}

\begin{abstract}
We apply the fidelity approach to study the topological superconductivity in a spin-orbit coupled nanowire system. The wire is modeled as a one layer lattice chain with Zeeman energy and spin-orbit coupling, which is in proximity to a multi layer superconductor. In particular, we study the effects of disorders and find that the fidelity susceptibility show multiple peaks. It is revealed that the major peak indicates the topological quantum phase transition, while other peaks signal the pining of the Majorana bound states by disorders.
\end{abstract}


\date{\today}
\end{frontmatter}

\section{Introduction}
Topological superconductors (TS) which host Majorana bound states (MBS) are under intensive study recently\cite{kitaev,read,ivanov,kanermp}. Superior to many other topological systems, MBSs are very useful non-Abelian quasiparticles. Spatially separated MBSs can build nonlocal topological qubits which is robust to environmental noises. More importantly, the braiding of these MBSs can rotate the topological qubits, acting effectively as a quantum gate\cite{ivanov,kanermp,alicea}. With these features, the TSs may constitute as an important building block for fault tolerant quantum computation\cite{kanermp,kitaev2,sarmarmp}.

TSs were originally introduced in spinless models\cite{kitaev,read}, which were thought unrealistic. Recently, it is realized that strong spin-orbit coupling in semiconductors may eliminate spin degree of freedom, making spinless superconductivity experimentally achievable\cite{fu}. After that, numerous proposals have appeared for realizing topological superconductivity in spin-orbit coupled systems by proximity to conventional s-wave superconductors, such as boundary of topological insulators\cite{fu}, spin-orbit coupled nanowires with Zeeman energy\cite{sau,lutchyn,oreg,beenakker,alicea2,law09}, carbon nanotubes\cite{loss12}, and ferromagnetic nanowires\cite{pergeprb}. Among them, the one dimensional nanowire systems are especially attractive due to recent experimental progresses\cite{kouwenhoven,rokhinson,heiblum,xu,rodrigo,harlingen,aguado1,pergescience}.

In one dimensional systems, it is understood that the effect of disorders might be significant. Disorders can modulate the differential conductance at the end of the wire, strongly influencing the signals from the Majorana bound states\cite{liujie}. For multiple band system, disorders will mix the subbands and enhance the zero conductance peak from the MBSs\cite{pientka}.
Meanwhile, disorders also have impacts on the
topological quantum phase transition (TQPT)\cite{beenakker-disorder}. Actually, with a combination of electron interaction, disorders may totally destroy the TQPT\cite{lobos}.
Considering the fact that the experimental results in recent nanowire systems are significantly different from the predictions in clean limit, more study on the disorder effect is important for the understanding of the topological superconductivity in realistic systems.

For studying the properties of TSs, one elementary method is the fidelity approach, which is a measure of difference between two ground state wave functions\cite{gureview}. When the system undergoes a dramatic change, the fidelity should present a peak on its susceptibility, making it an ideal marker for studying the ground state properties\cite{gufidelity}. Fidelity approach is a general method to depict all types of QPTs, while it is especially suitable for treating TQPT where the local order parameters are missing. Application of fidelity approach in several models with TQPTs has obtained satisfactory results\cite{gureview}.
Fidelity approach has also been used to study the toy model of Kitaev wire with disorders. It was shown that the peak in fidelity susceptibility is able to mark the TQPT in the toy model\cite{wangfidelitynew}. With these results, it is natural to introduce fidelity approach to study the realistic nanowire system.

In this work, we study the fidelity susceptibility in one dimensional systems with spin-orbit coupling, Zeeman energy, and proximity induced superconducting gap. We numerically solve the Bogoliubov-de Gennes (BdG) equations to obtain the ground state wave functions of the system and calculate the fidelity susceptibility. In the clean limit, we find that one single peak appears at the TQPT point predicted by previous analytic result, which verifies the ability of fidelity approach to mark the TQPT. More interestingly, we find much richer phenomenon in disordered systems. Multiple peaks in fidelity susceptibility appear for disordered systems, showing that fidelity approach may signal more information other than TQPT.
In order to understand these peaks better, we calculate the zero energy local density of states (LDOS) on the wire. Comparing the results, we find that one of the peaks in fidelity susceptibility signals the TQPT, where the zero energy MBSs emerge; other peaks signal the movement of the MBS from the end to the inner part of the wire, which is a pining effect of the disorder. We explore different types of disorders and more realistic systems with substrates, and find that this disorder pining effect is quite universal in all systems.

This paper is organized as follows: the fidelity approach we used to study the system is presented in section 2; the results for the clean and disordered one dimensional systems are shown in section 3. In section 4 we discuss a more realistic model by considering the effect of the substrate.

\section{Fidelity approach}
Recent experiments report the discovery of zero-bias peaks in a spin-orbit coupled nanowire in proximity to conventional s-wave superconductivity\cite{kouwenhoven}, making it a promising candidate for topological superconductor. We use a tight-binding model to describe the system,
\begin{eqnarray}
H=&& \sum\limits_{\langle \bm{r},\bm{r'}\rangle\sigma} t_{\bm{r},\bm{r}'} c^\dagger_{\bm{r}\sigma}c_{\bm{r'}\sigma}- \mu\sum\limits_{\bm{r}\sigma} c^\dagger_{\bm{r}\sigma}c_{\bm{r}\sigma}
+\frac{\alpha}{2} \sum\limits_{\bm{r}\bm{\delta}\sigma\sigma'}c^\dagger_{\bm{r}+\bm{\delta},\sigma}(i\tau_y)_{\sigma\sigma'}c_{\bm{r}\sigma'}
\\&&+\sum\limits_{\bm{r}\sigma\sigma'}c^\dagger_{\bm{r}\sigma}[V_x\tau_x+V_y\tau_y]_{\sigma\sigma'}c_{\bm{r}\sigma'}
+ \Delta \sum\limits_{\bm{r}} c^\dagger_{\bm{r}\uparrow}c^\dagger_{\bm{r}\downarrow}
+\sum_{\bm{r}\sigma}V_{dis}(\bm{r})c^\dagger_{\bm{r}\sigma}c_{\bm{r}\sigma} + h.c. \nonumber
\end{eqnarray}
where $\bm{r}$ is a two dimensional vector indicating the sites on a $W \times L$ lattice with $W$ the layer width and $L$ the length, $\sigma$ is the electron spin, $\tau_{x,y}$ are pauli matrices,  $\bm{\delta}$ is the unit vector in the $x$ or $y$ direction, $t$ is the hopping integral, $\mu$ is the chemical potential, $\alpha$ is the spin-orbit coupling strength, $V_{x,y}$ are the Zeeman energy from the magnetic field in x and y directions respectively, $\Delta$ is the s-wave pairing amplitude from proximity effect, and $V_{dis}({\bm r})$ is the strength distribution function of the disorders. This Hamiltonian includes all important ingredients for topological superconductivity.

 Numerically solving the BdG equations associated with Eq. (1), we can diagonalize the Hamiltonian to obtain the
eigenenergies $E_n$ and eigenstates $\phi_{n}({\bf r}) = ({\rm u}_{\uparrow, n,{\bf r}},{\rm u}_{\downarrow, n,{\bf r}},{\rm v}_{\uparrow,n,{\bf r}},{\rm v}_{\downarrow,n,{\bf r}})$, where $n$ denotes the quantum number of the energy levels.
The ground state wave function can be obtained through a combination of wave functions of all negative energy quasi-particle states, which is in a Slater determinant form\cite{wangfidelitynew},
\begin{eqnarray}
|\Phi({\bf r}_1,\cdots,{\bf r}_N) \rangle=
 \frac{1}{\sqrt{n!}}\left|
 \begin{array}{ccc}
\phi_1({{\bf r}_1})&\cdots&\phi_1({{\bf r}_N})\\
 \vdots&\ddots&\vdots\\
  \phi_n({{\bf r}_1})&\cdots&\phi_n({{\bf r}_N})
\end{array}
  \right|,
\end{eqnarray}
where $N$ denotes the number of quasi-particle states with negative energy.

\begin{figure}[ht]
\begin{center}
\includegraphics[clip=true, scale=0.5]{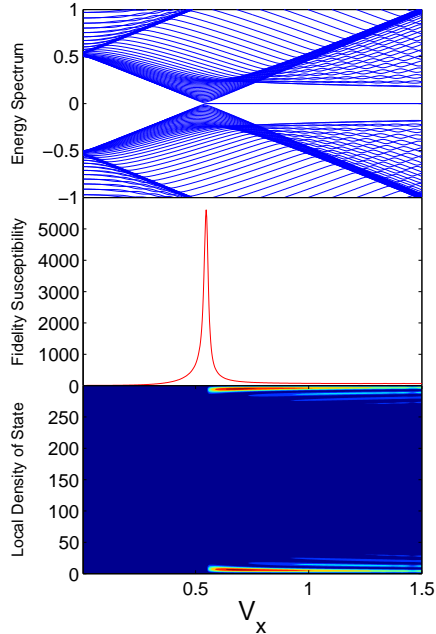}
\caption{Three important physical properties associated with topological phase transition in the nanowire, with driving field $V_x$: the energy spectrum (upper panel), the fidelity susceptibility (middle panel), the zero energy LDOS (lower panel, where the brighter zone indicates the higher LDOS). The parameters for the numerical calculation are $t_{i,j}=-11$ when $|i-j|=1$, $t_{i,j}=22$ when $i=j$, $\mu=0.2$, $\alpha=1.34$, $V_y=0$ and $\Delta=0.5$.}
\end{center}
\end{figure}

The fidelity approach has been used to identify the TQPT in many systems, and proven to be useful in studying the TQPT point. In previous research, fidelity susceptibility has also been introduced to study the Kitaev Hamiltonian, which is a toy model for topological superconductivity. It was shown that fidelity susceptibility is able to signal the TQPT point even under the presence of disorders. Now we want to expand this approach to study the realistic model.
In the fidelity approach, we are interested in the sudden changes of the ground states, thus a driving parameter is needed. For current systems, we adopt the Zeeman energy $V_x$, which is convenient since it directly relates to the splitting of the spin degree of freedom. When the Zeeman energy is modulated,
we define fidelity as the likeness of the ground-states,
\begin{equation}
F=\langle\Phi({V_x}) | \Phi(V_x+ \delta V_x)\rangle,
\end{equation}
where $\delta V_x$ is a small variation of the Zeeman energy which is hand adopted. It is obvious that the fidelity is strongly influenced by this choice of step length. A better method for signaling ground state change would be the fidelity susceptibility, which is the differentiation of the fidelity\cite{wangfidelitynew},
\begin{eqnarray}
\chi_F&&=  2 \lim_{\delta V_x \rightarrow 0} \frac{1-|\langle \Phi({V_x})|\Phi(V_x+ \delta V_x) \rangle|^2}{\delta V_x^2}.
\end{eqnarray}
With these expressions, the fidelity susceptibility is readily obtained once the BdG equations are solved numerically.

\section{Results for one dimensional system}
Now we apply the fidelity approach to study the realistic model described by Eq. (1). As a bench mark test, it would be helpful to start our calculations from the clean system. For this simple model, analytic results have been obtained. It has been shown that energy gap of the system would close at $V_x=\sqrt{\mu^2+\Delta^2}$. Since the topology of the system is protected by the energy gap, this particular point would be the phase transition point if the TQPT exists. On the other hand, it has been shown that the MBSs appear at one side of the point, and disappears on the other side. It is natural to conclude that a TQPT happens with the appearance of the MBS as a signature of the topological nontrivial phase.

In this section, we concentrate on the one dimensional system. We numerically solve the BdG equations for the single layer case $W=1$ in Hamiltonian Eq. (1) without disorders, where the length is chosen to be $L=300$.
We obtain the fidelity susceptibility as shown in the middle panel of Fig. 1. We find that one single peak appears which should indicate the expected TQPT point predicted by the analytic results.
We check the position of the peak, finding it to be around $V_x \approx 0.538$, which agrees well with the analytic results. As a comparison, we also demonstrate the energy spectrum and the zero energy LDOS from the numerical calculations in the upper and lower panels of Fig. 1, respectively. We realize that the energy gap extracted from the energy spectrum closes exactly at the same position, while the zero energy bound states at the end of the wire also appears at that point. These consistent numerical results from fidelity susceptibility, energy spectrums, and LDOS verify the story from the analytic arguments, that is, a TQPT should happen before the MBSs appear at the end of the wire. Our numerical results prove that the fidelity approach is useful for studying the MBSs in this realistic model.

\begin{figure}[tb]
  \centering
   \includegraphics[clip, scale=0.5]{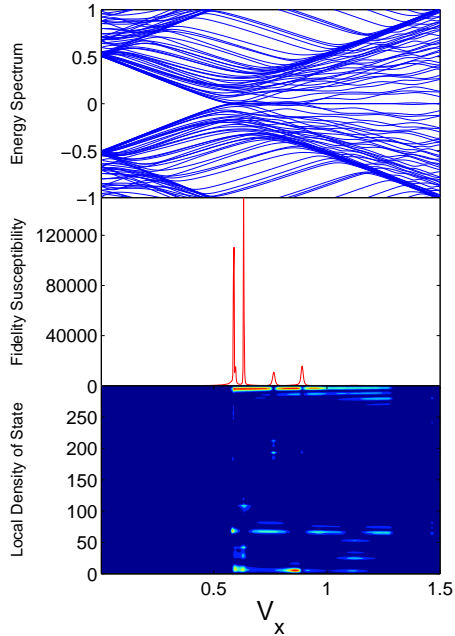}
   \caption{Same as in Fig. 1 for disordered system, with the disorder parameter $\epsilon=4$.}
\end{figure}

A more interesting problem in the current topological systems is the influence of disorders. Since disorders are inevitable in realistic experimental devices, any topological arguments would be meaningless if it cannot survive moderate disorders. For current one dimensional nanowire system, the problem is more important since we know that the impact of disorders gets stronger for lower dimension.
Previous researches have shown that disorders may variate the coupling between end MBS, and may also produce zero-bias conductance peak at the trivial phase of topological superconductors. These results suggest that disorder may significatively change the picture of topological superconductivity. Other researches, however, shows that disorders do not qualitatively influence the TQPT or MBSs. To clarify the problem, we apply the fidelity approach to investigate the realistic model with disorders.

We solve the Hamiltonian Eq. (1) with the disorder distribution function $V_{dis}({\bm r})$. We consider randomly distributed on-site disorder, with strength chosen independently for different sites. For each disordered site, $V_{dis}$ varies according to a Gaussian distribution function, $f(V,0,\epsilon)=\frac{1}{\epsilon\sqrt{2\pi}}e^{-\frac{V^2}{2\epsilon^2}}$.
We solve the BdG equations for $10\%$ disordered sites, and show the results of the fidelity susceptibility in the middle panel of Fig. 2. Intriguingly, we find that there are multiple peaks in fidelity susceptibility for the disordered systems, as compared to the single peak in the clean system. Obviously, not all of these peaks signify the TQPT. A better understanding of these peaks is demanded, since the ground state wave function must have dramatically changed at these points. For this purpose, we present the energy spectrums in the upper panel of Fig. 2 as comparison. We find that, as $V_x$ increases from zero, the energy gap of the system gradually shrinks and finally closes at the point coincide with the first peaks in fidelity susceptibility. With this observation, we can claim that the first peak in fidelity susceptibility really signals the TQPT. However, after the TQPT, the energy spectrum becomes obscured by the excited states from disorders, and little information can be obtained for the other peaks.

For further understanding, we calculate the zero energy LDOS of the system as shown in the lower panel of Fig. 2. We find that zero energy LDOS appears after the first fidelity susceptibility peak, providing consistent results for its identification of TQPT. More interestingly, we find that
the location of the zero energy states changes after the second peak in fidelity susceptibility, it moves from the end of wire to the inner part of the wire. Since these zero energy states are expected as MBSs, this process actually describes the pinning of the MBSs by disorders.
For the third and fourth peak in the fidelity susceptibility, we find the similar phenomena. That is, the position of the MBSs extracted from LDOS changes dramatically at the peak positions. From this strong correlation, it is natural to claim that the peaks in the fidelity susceptibility not only  signal the TQPT, but also describe the pinning of the MBSs by disorders. Our results reveal that the fidelity approach is a powerful method to investigate the topological superconductors. The peaks in Fidelity susceptibility can provide the information for both the TQPT and the behavior of MBSs.


\section{Results with Substrate}
The possible topological superconducting systems investigated in recent experiments are mostly hybrid systems, with a spin-orbit coupled nanowire placed upon a conventional s-wave superconducting substrate. It has been shown that the substrate may significantly influence the MBSs on the wire. Meanwhile, it is also important to notice that even for clean system without doping impurities on either the wire or the substrate, the inhomogeneous contact between them will inevitably introduce some disorders in electron hoping integral. In order to study this type of disorder, we consider a specific Hamiltonian with the wire modeled by a single layer chain with spin-orbit coupling and the substrate modeled by multi-layers with conventional s-wave superconductivity,
\begin{eqnarray}
H&&=\sum\limits_{\langle i,j\rangle,\sigma}t_{1,i,j}c^\dagger_{i\sigma}c_{j\sigma}-\mu\sum\limits_{i,\sigma} c^\dagger_{i\sigma}c_{i\sigma}+\frac{\alpha}{2}\sum\limits_{i,\sigma\sigma'}c^\dagger_{i+1\sigma}(i\tau_y)_{\sigma\sigma'}c_{i\sigma'}\nonumber
\\&&+\sum\limits_{i,\sigma}c^\dagger_{i\sigma}[V_x\tau_x+V_y\tau_y]_{\sigma\sigma'}c_{i\sigma'}\nonumber
+\sum\limits_{\langle l,m\rangle,\sigma}t_{2,l,m}a^\dagger_{l\sigma}a_{m\sigma}-\mu\sum\limits_{l,\sigma}a^\dagger_{l\sigma}a_{l\sigma}\\&& +\Delta\sum\limits_l a^\dagger_{l\uparrow}a^\dagger_{l\downarrow}+t_{12}\sum\limits_{\langle i,l\rangle\sigma}a^\dagger_{i\sigma}c_{i\sigma} + h.c,
\end{eqnarray}
where $c^\dagger$ and $a^\dagger$ is the electron creation operator of the one layer wire and multi-layer substrate respectively, $t_{1,2}$ are the hoping integrals.
\begin{figure}[hbt]
  \centering
   \includegraphics[clip, scale=0.5]{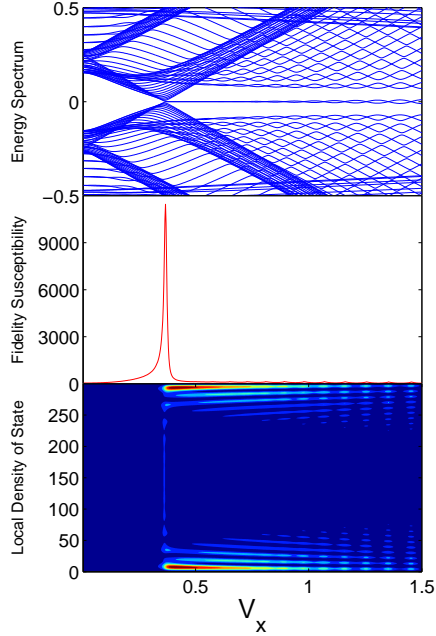}
   \caption{Same as for Fig. 1 with two more layers of substrate. The parameter in numerical calculations are $t_{2,l,m}=-6$ when $|l-m|=1$, $t_{2,l,m}=12$ when $l=m$, $t_{12}=4$, and other parameters the same as in Fig. 1.}
\end{figure}

\begin{figure}[hbt]
  \centering
   \includegraphics[clip, scale=0.5]{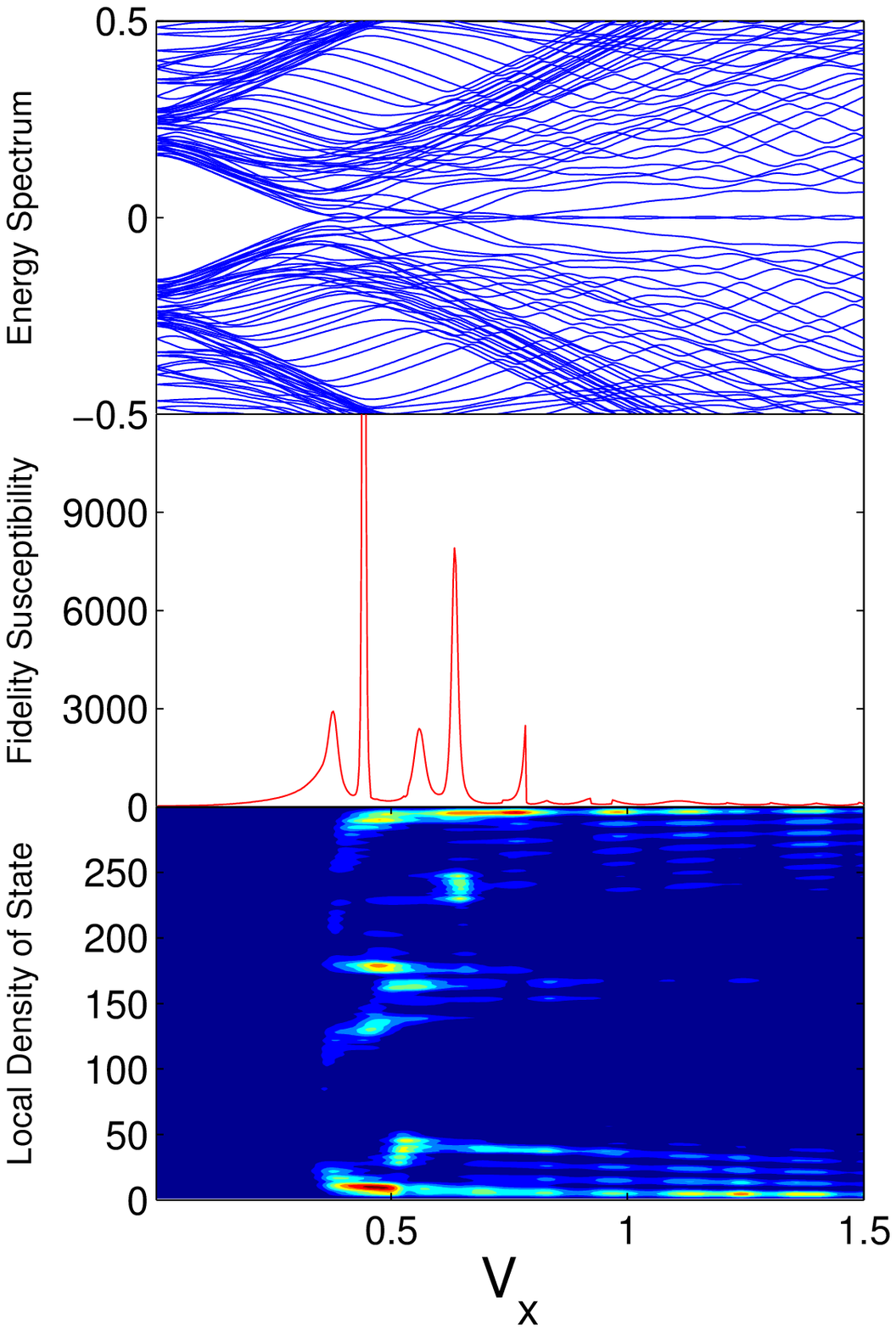}
   \caption{Same as in Fig. 3 for disordered system, with the disorder parameter $\epsilon=4$.}
\end{figure}

Without disorders, the Hamiltonian Eq. (5) should lead to similar results as in the one dimensional model. In Fig. 3, we show the results for the single layer nanowire in contact with two layers of superconducting substrate. We find that the energy spectrum and fidelity susceptibility indeed resembles the result of previous one dimensional model, with the energy gap closing and the single peak in fidelity susceptibility indicate the TQPT. As for the zero energy LDOS, We notice that the wave functions of the MBSs are broader, comparing with the results obtained in the one dimensional model. Moreover, our numerical results are in accord with the theoretical prediction that the substrate results in larger interaction between end MBSs\cite{zyuzin}.

The disorders in the system may come from two origins, the impurity disorder due to the doping of ions and vacancies, and the coupling disorder from the inhomogeneous contact between the wire and the substrate. While the former type of disorder is largely reduced by the material growing technology, the latter type is inevitable in realistic systems. We show the results with 10\% of Gaussian type disorders on $t_{12}$ in Fig. 4, and find that the MBS pining effect also appears, as clearly marked in the fidelity susceptibility and the local density of states. We also calculated the results with on-site disorders on substrate, and find very similar results of the MBS pining. That is, multiple peaks in fidelity susceptibility appear, with one of them indicating the TQPT and others describing the MBS pining. Our calculations with a combination of fidelity susceptibility and local density of states show that the MBS pining is a general phenomenon for different models and disorders.


\section{Conclusion}
In summary, we study the effects of disorders in topological superconductors within the fidelity approach. We adopted the realistic models with the spin-orbit coupled nanowire in proximity to a conventional superconducting substrate, and showed that a peak in fidelity susceptibility signals the topological quantum phase transition in the pure system. With the introduction of  moderate disorders, we noticed that the location of the Majorana bound states can shift to the inner part of the wire from the ends of the wire, which is a pining effect of disorder. We systematically studied the different models and disorders, and found that the pining effect is a general behavior in disordered topological superconductors, which can be measured by experiments.

\section*{Acknowledgement}
This work was supported by NSFC-11304400, SRFDP-20130171120015, 985 Project of Sun Yat-Sen University.
D.X.Y. is supported by the National Basic Research Program of China (2012CB821400), NSFC-11074310, NSFC-11275279, Specialized Research Fund for the Doctoral Program of Higher Education (20110171110026), Fundamental Research Funds for the Central Universities of China, and NCET-11-0547.





\bibliographystyle{elsarticle-num}
\bibliography{<your-bib-database>}



\end{document}